%% file: main.tex
\documentclass[10pt,journal,compsoc]{IEEEtran}
\usepackage{microtype}                 % use micro-typography (slightly more compact, better to read)
\PassOptionsToPackage{warn}{textcomp}  % to address font issues with \textrightarrow
\usepackage{textcomp}                  % use better special symbols
\usepackage{mathptmx}                  % use matching math font
\usepackage{times}                     % we use Times as the main font
\usepackage{tabu}                      % only used for the table example
\usepackage{booktabs}                  % only used for the table example
\usepackage{cite}

\usepackage{paralist}
\usepackage{amssymb}
\usepackage{float}
\usepackage{placeins}
\usepackage{scrextend}
\usepackage{graphicx}
\usepackage[table,xcdraw]{xcolor}
\usepackage{makecell}
\usepackage{hyperref}
\usepackage{enumerate}

\usepackage{xspace,xpunctuate}
\newcommand{\ie}{{i.e.,}\xspace}
\newcommand{\eg}{{e.g.,}\xspace}

\definecolor{indicator}{RGB}{177, 47, 31}
\definecolor{explanation}{RGB}{60, 103, 143}
\definecolor{knowledge}{RGB}{91, 157, 59}

\definecolor{salmon}{rgb}{1, 0.55, 0.42}

\definecolor{mygreen}{RGB}{34,139,34}
\definecolor{myblue}{RGB}{48, 85, 152}
\definecolor{myorange}{RGB}{204, 112, 0}
\definecolor{myred}{RGB}{190, 0, 0}

\newcommand{\rone}[1]{\textcolor{black}{#1}}
\newcommand{\rtwo}[1]{\textcolor{black}{#1}}
\newcommand{\rthree}[1]{\textcolor{black}{#1}}

\ifCLASSINFOpdf
\else
\fi
\begin{document}
\title{\title{Explaining with Examples: Lessons Learned from Crowdsourced Introductory Description of Information Visualizations}}
%
%
% author names and IEEE memberships
% note positions of commas and nonbreaking spaces ( ~ ) LaTeX will not break
% a structure at a ~ so this keeps an author's name from being broken across
% two lines.
% use \thanks{} to gain access to the first footnote area
% a separate \thanks must be used for each paragraph as LaTeX2e's \thanks
% was not built to handle multiple paragraphs
%
%
%\IEEEcompsocitemizethanks is a special \thanks that produces the bulleted
% lists the Computer Society journals use for "first footnote" author
% affiliations. Use \IEEEcompsocthanksitem which works much like \item
% for each affiliation group. When not in compsoc mode,
% \IEEEcompsocitemizethanks becomes like \thanks and
% \IEEEcompsocthanksitem becomes a line break with idention. This
% facilitates dual compilation, although admittedly the differences in the
% desired content of \author between the different types of papers makes a
% one-size-fits-all approach a daunting prospect. For instance, compsoc 
% journal papers have the author affiliations above the "Manuscript
% received ..."  text while in non-compsoc journals this is reversed. Sigh.

\author{Leni~Yang,~\IEEEmembership{}
        Cindy~Xiong,~\IEEEmembership{}
        Jason~K.~Wong,~\IEEEmembership{}
        Aoyu~Wu~\IEEEmembership{}
        and Huamin~Qu~\IEEEmembership{}% <-this % stops a space
\IEEEcompsocitemizethanks{\IEEEcompsocthanksitem L. Yang, J. Wong, A.Wu and H. Qu are with the Hong Kong University of Science and Technology. Email: \{lyangbb, kkwongar, awuac, huamin\}@cse.ust.hk.
\IEEEcompsocthanksitem C. Xiong is with University of Massachusetts Amherst. Email: cindy.xiong@cs.umass.edu}% <-this % stops an unwanted space
\thanks{}}

% note the % following the last \IEEEmembership and also \thanks - 
% these prevent an unwanted space from occurring between the last author name
% and the end of the author line. i.e., if you had this:
% 
% \author{....lastname \thanks{...} \thanks{...} }
%                     ^------------^------------^----Do not want these spaces!
%
% a space would be appended to the last name and could cause every name on that
% line to be shifted left slightly. This is one of those "LaTeX things". For
% instance, "\textbf{A} \textbf{B}" will typeset as "A B" not "AB". To get
% "AB" then you have to do: "\textbf{A}\textbf{B}"
% \thanks is no different in this regard, so shield the last } of each \thanks
% that ends a line with a % and do not let a space in before the next \thanks.
% Spaces after \IEEEmembership other than the last one are OK (and needed) as
% you are supposed to have spaces between the names. For what it is worth,
% this is a minor point as most people would not even notice if the said evil
% space somehow managed to creep in.

% The paper headers
\markboth{Authors' preprint}%
{Shell \MakeLowercase{\textit{et al.}}: Bare Demo of IEEEtran.cls for Computer Society Journals}
% The only time the second header will appear is for the odd numbered pages
% after the title page when using the twoside option.
% 
% *** Note that you probably will NOT want to include the author's ***
% *** name in the headers of peer review papers.                   ***
% You can use \ifCLASSOPTIONpeerreview for conditional compilation here if
% you desire.

% The publisher's ID mark at the bottom of the page is less important with
% Computer Society journal papers as those publications place the marks
% outside of the main text columns and, therefore, unlike regular IEEE
% journals, the available text space is not reduced by their presence.
% If you want to put a publisher's ID mark on the page you can do it like
% this:
%\IEEEpubid{0000--0000/00\$00.00~\copyright~2015 IEEE}
% or like this to get the Computer Society new two part style.
%\IEEEpubid{\makebox[\columnwidth]{\hfill 0000--0000/00/\$00.00~\copyright~2015 IEEE}%
%\hspace{\columnsep}\makebox[\columnwidth]{Published by the IEEE Computer Society\hfill}}
% Remember, if you use this you must call \IEEEpubidadjcol in the second
% column for its text to clear the IEEEpubid mark (Computer Society jorunal
% papers don't need this extra clearance.)

% use for special paper notices
%\IEEEspecialpapernotice{(Invited Paper)}

% for Computer Society papers, we must declare the abstract and index terms
% PRIOR to the title within the \IEEEtitleabstractindextext IEEEtran
% command as these need to go into the title area created by \maketitle.
% As a general rule, do not put math, special symbols or citations
% in the abstract or keywords.
\IEEEtitleabstractindextext{%
\begin{abstract}
Data visualizations have been increasingly used in oral presentations to communicate data patterns to the general public.
Clear verbal introductions of visualizations to explain how to interpret the visually encoded information are essential to convey the takeaways and avoid misunderstandings. 
We contribute a series of studies to investigate how to effectively introduce visualizations to the audience with varying degrees of visualization literacy.
We begin with understanding how people are introducing visualizations.
We crowdsource 110 introductions of visualizations and categorize them based on their content and structures.
From these crowdsourced introductions, we identify different introduction strategies and generate a set of introductions for evaluation.
We conduct experiments to systematically compare the effectiveness of different introduction strategies across four visualizations with 1,080 participants. 
We find that introductions explaining visual encodings with concrete examples are the most effective.
Our study provides both qualitative and quantitative insights into how to construct effective verbal introductions of visualizations in presentations, inspiring further research in data storytelling.
\end{abstract}

% Note that keywords are not normally used for peerreview papers.
}

% make the title area
\maketitle

% To allow for easy dual compilation without having to reenter the
% abstract/keywords data, the \IEEEtitleabstractindextext text will
% not be used in maketitle, but will appear (i.e., to be "transported")
% here as \IEEEdisplaynontitleabstractindextext when the compsoc 
% or transmag modes are not selected <OR> if conference mode is selected 
% - because all conference papers position the abstract like regular
% papers do.
\IEEEdisplaynontitleabstractindextext
% \IEEEdisplaynontitleabstractindextext has no effect when using
% compsoc or transmag under a non-conference mode.

% For peer review papers, you can put extra information on the cover
% page as needed:
% \ifCLASSOPTIONpeerreview
% \begin{center} \bfseries EDICS Category: 3-BBND \end{center}
% \fi
%
% For peerreview papers, this IEEEtran command inserts a page break and
% creates the second title. It will be ignored for other modes.
\IEEEpeerreviewmaketitle

\input{section/1-introduction}
\input{section/2-related-work}
\input{section/3-pilot-study}

\input{section/4-study1}
\input{section/5-study2}
\input{section/6-discussion}
\input{section/7-conclusion}

% if have a single appendix:
%\appendix[Proof of the Zonklar Equations]
% or
%\appendix  % for no appendix heading
% do not use \section anymore after \appendix, only \section*
% is possibly needed

% use appendices with more than one appendix
% then use \section to start each appendix
% you must declare a \section before using any
% \subsection or using \label (\appendices by itself
% starts a section numbered zero.)
%

% \appendices
% \section{Proof of the First Zonklar Equation}
% Appendix one text goes here.

% % you can choose not to have a title for an appendix
% % if you want by leaving the argument blank
% \section{}
% Appendix two text goes here.

% use section* for acknowledgment
\ifCLASSOPTIONcompsoc
  % The Computer Society usually uses the plural form
  \section*{Acknowledgments}
\else
  % regular IEEE prefers the singular form
  \section*{Acknowledgment}
\fi

We thank Meghna Jain, Chase Stokes, and Madison Tyrcha for their generous help on coding the data and their precious feedback. We also would like to thank anonymous reviewers for their constructive comments. This research was supported in part by Hong Kong Theme-based Research Scheme Grant T41-709/17N.

% Can use something like this to put references on a page
% by themselves when using endfloat and the captionsoff option.
\ifCLASSOPTIONcaptionsoff
  \newpage
\fi

% trigger a \newpage just before the given reference
% number - used to balance the columns on the last page
% adjust value as needed - may need to be readjusted if
% the document is modified later
%\IEEEtriggeratref{8}
% The "triggered" command can be changed if desired:
%\IEEEtriggercmd{\enlargethispage{-5in}}

% references section

\bibliographystyle{IEEEtran}
\bibliography{bibliography}

% \bibitem{IEEEhowto:kopka}
% H.~Kopka and P.~W. Daly, \emph{A Guide to \LaTeX}, 3rd~ed.\hskip 1em plus
%   0.5em minus 0.4em\relax Harlow, England: Addison-Wesley, 1999.

% \end{thebibliography}

% biography section
% 
% If you have an EPS/PDF photo (graphicx package needed) extra braces are
% needed around the contents of the optional argument to biography to prevent
% the LaTeX parser from getting confused when it sees the complicated
% \includegraphics command within an optional argument. (You could create
% your own custom macro containing the \includegraphics command to make things
% simpler here.)
%\begin{IEEEbiography}[{\includegraphics[width=1in,height=1.25in,clip,keepaspectratio]{mshell}}]{Michael Shell}
% or if you just want to reserve a space for a photo:

% \vspace{-10mm}
\begin{IEEEbiography}[{\includegraphics[width=1in,height=1.25in,clip,keepaspectratio]{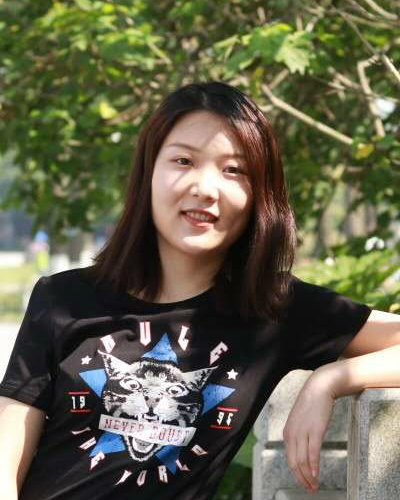}}]{Leni Yang} 
is a Ph.D. student in the Department of Computer Science and Engineering (CSE) at the Hong Kong University of Science and Technology (HKUST). She received her a B.E. in MIS from University of Electronic Science and Technology of China. Her research interests lie in data visualization and human-computer interaction, with focuses on data-driven storytelling. 
\end{IEEEbiography}

\begin{IEEEbiography}[{\includegraphics[width=1in,height=1.25in,clip,keepaspectratio]{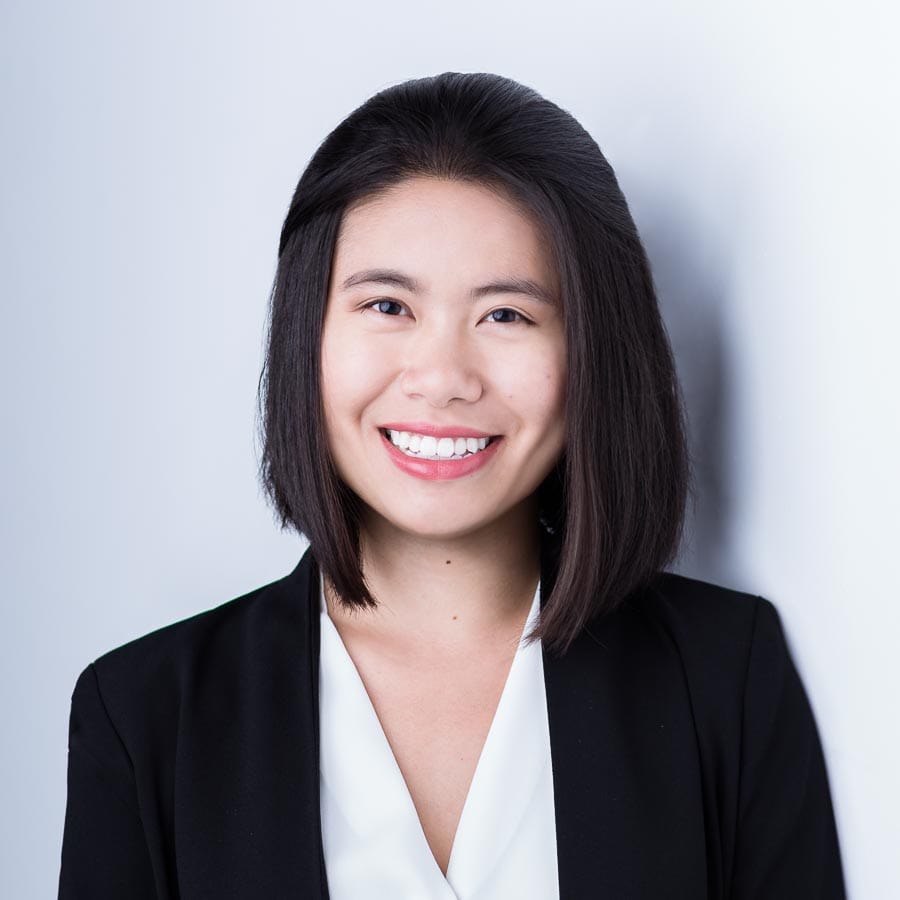}}]{Cindy Xiong} 
is an Assistant Professor in the College of Information and Computer Sciences at UMass Amherst. Her research program combines visual perception, cognition, and data visualization. By investigating how humans perceive, interpret, and make decisions from visualized data, she aims to improve visualization design, data storytelling, and data-driven decision-making.
\end{IEEEbiography}

\begin{IEEEbiography}[{\includegraphics[width=1in,height=1.25in,clip,keepaspectratio]{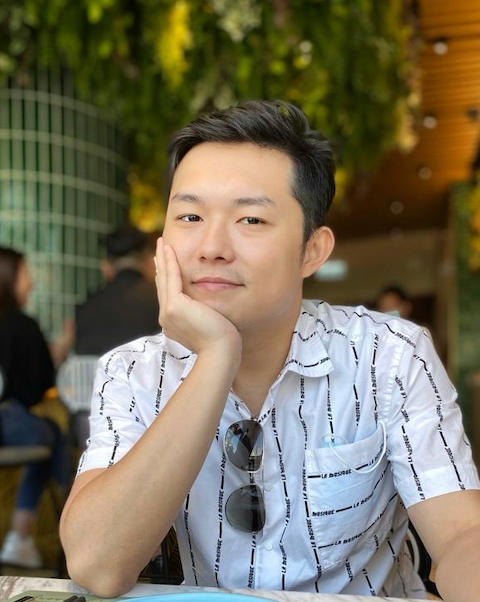}}]{Jason K. Wong} 
is a Ph.D. student in the Department of Computer Science and Engineering (CSE) at the Hong Kong University of Science and Technology (HKUST). He received his a B.E. in CSE from HKUST. His main research interests are in data visualization, visual analytics and data mining. He is passionate about their applications in financial data and FinTech.
\end{IEEEbiography}

\begin{IEEEbiography}[{\includegraphics[width=1in,height=1.25in,clip,keepaspectratio]{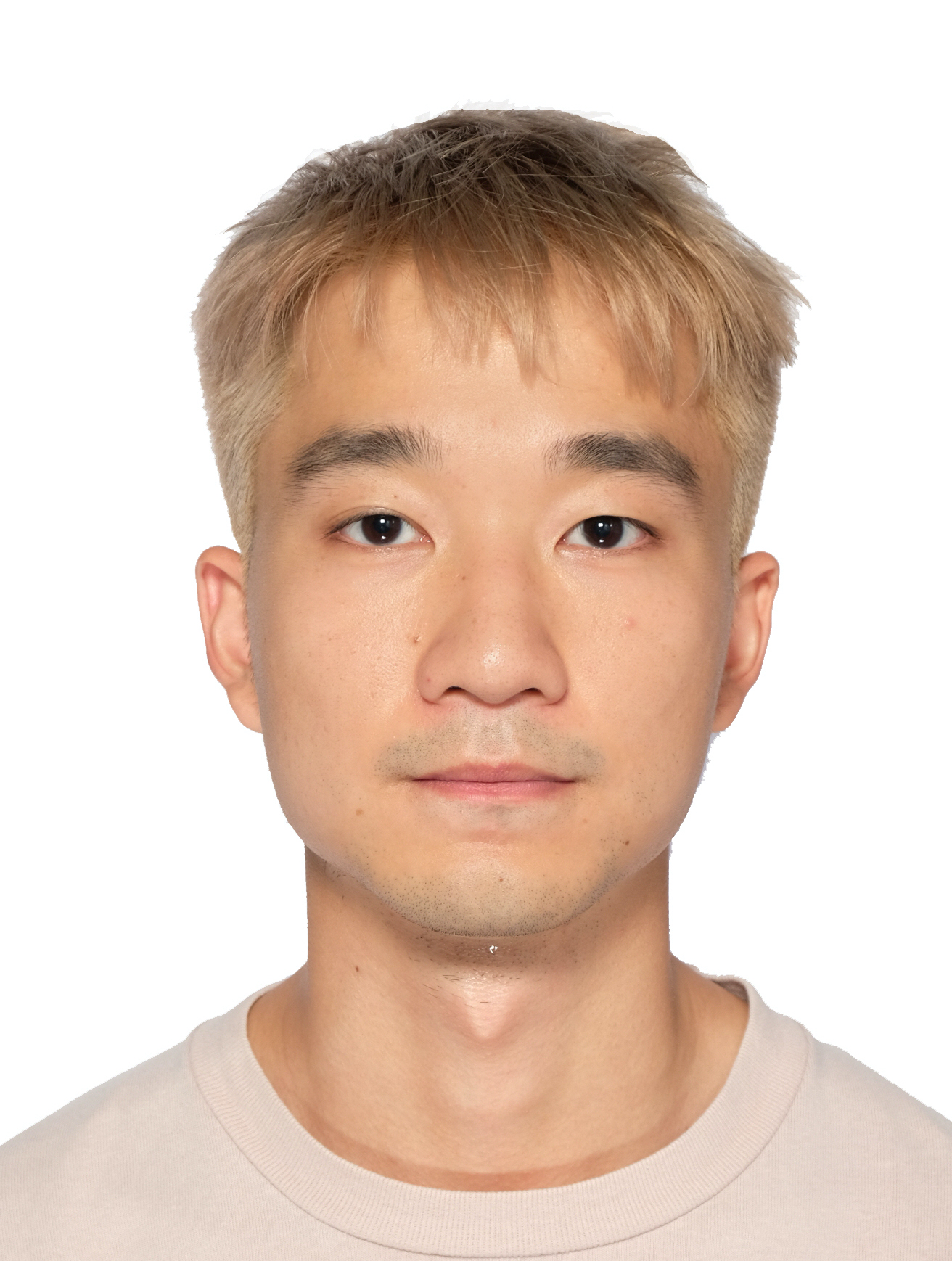}}]{Aoyu Wu} 
is a Ph.D. student in the Department of Computer Science and Engineering at the Hong Kong University of Science and Technology (HKUST). He received his B.E. and M.E. degrees from HKUST. His research interests include data visualization and human-computer interaction.
For more details, please refer to \url{http://awuac.student.ust.hk/.}
\end{IEEEbiography}

\begin{IEEEbiography}[{\includegraphics[width=1in,height=1.25in,clip,keepaspectratio]{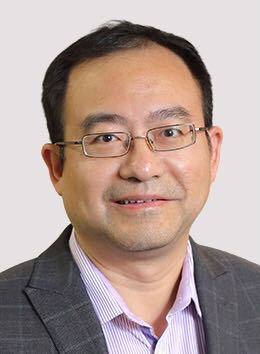}}]{Huamin Qu} 
is a professor in the Department of Computer Science and Engineering (CSE) at the Hong Kong University of Science and Technology (HKUST) and also the director of the interdisciplinary program office (IPO) of HKUST. He obtained a BS in Mathematics from Xi'an Jiaotong University, China, an MS and a PhD in Computer Science from the Stony Brook University. His main research interests are in visualization and human-computer interaction, with focuses on urban informatics, social network analysis, E-learning, text visualization, and explainable artificial intelligence (XAI).
\end{IEEEbiography}

% insert where needed to balance the two columns on the last page with
% biographies
%\newpage

% You can push biographies down or up by placing
% a \vfill before or after them. The appropriate
% use of \vfill depends on what kind of text is
% on the last page and whether or not the columns
% are being equalized.

%\vfill

% Can be used to pull up biographies so that the bottom of the last one
% is flush with the other column.
%\enlargethispage{-5in}

% that's all folks
\end{document}

%% file: section/1-introduction.tex
\IEEEraisesectionheading{\section{Introduction}\label{sec:introduction}}

\IEEEPARstart{D}{ata} visualizations have become increasingly popular with the ongoing data democratization. Visualizing data improves the efficiency and effectiveness of delivering data facts and communicating data stories to the general public. As such, they are often incorporated in presentations and talks in a wide variety of formal and informal settings~\cite{kosara2013storytelling}.
In these circumstances, an effective introduction is necessary to ensure communication clarity and audience engagement~\cite{lee2015people,kosara2016presentation}. An effective introduction should help the audience understand how information is visually encoded in the visualization. 
However, presenters who are familiar with their visualizations might suffer from the ``curse of expertise''~\cite{xiong2019curse}. They tend to assume that the audience is also familiar with the visualizations and give an inadequate introduction. As a result, their audience would be confused or overwhelmed. 

Visualization research has proposed various forms of guidance on how to read a chart, such as animation~\cite{ruchikachorn2015learning}, cheat sheets~\cite{wang2020cheat}, and text-plus-questions~\cite{tanahashi2016study}. However, little research has investigated how to introduce a visualization in presentations effectively. 
Research in education has identified various factors that contribute to effective verbal explanation, such as giving topic sentences~\cite{lorch1985topic,lorch1996effects,schwarz1981text} and providing concrete examples~\cite{shimoda1993effects,sadoski2000engaging,sadoski2001resolving}. However, little has been known about their effects in the context of explaining data visualizations, which are considerably more challenging.
% , especially for novel, unfamiliar visualization types. 
To address this gap, we aim to investigate effective approaches to giving verbal introductions of visualizations to the general public. \rthree{Our study focuses on verbal introductions as the auditory channel is the primary source from which the audience obtains and understands the content of an oral presentation~\cite{bergen2005attention}}.

We began with a qualitative investigation into the current state of how people were introducing visualizations. Specifically, we conducted a crowdsourced experiment where we showed a visualization to participants with varying visualization literacy and prompted them to write an introductory description for it. We collected introductions from 110 participants and obtained an inclusive corpus of introductions. We documented interesting and common writing techniques people incorporated in their introductions, such as including topic sentences and providing examples to show what each visual encoding channel represents. 
% We found that some strategies are similar to popular practices in education, while others are specified for introductions of visualizations. 
Then, we conducted follow-up experiments to investigate which introduction strategy best helped visualization comprehension. We recruited participants to listen to an audio recording of our introductions while looking at a visualization, mimicking an oral presentation scenario~\cite{kong2019understanding,kong2017internal}. 
The participants completed a visualization comprehension test based on tasks introduced by Lee et al.~\cite{lee2016vlat} afterwards. 
We found that the most effective introductions explained the meaning of visual encoding channels with concrete examples. Interestingly, including topic sentences, which had been shown to work well to promote understanding in education research~\cite{lorch1985topic,lorch1996effects,schwarz1981text}, did not have a significant effect on increasing visualization comprehension. Finally, we discussed the potential reasons behind our results and presented implications for composing effective verbal introductions for visualizations and pointed to future research opportunities in this area. All the study materials are available at the link\footnote{\url{https://osf.io/asvt6/?view_only=559eb881e3f640aa984d0ca8ad593d0a}}. 

In summary, the major contributions of this work include:
\begin{itemize}
\item A survey of how the general public introduces visualizations
\item A series of experiments investigating what makes an introduction effective
\item A set of implications of how to introduce visualizations to the general public
\end{itemize}

%% file: section/2-related-work.tex
\section{Related Work}

\begin{figure*}[t]
  \centering
  \includegraphics[width=1\textwidth]{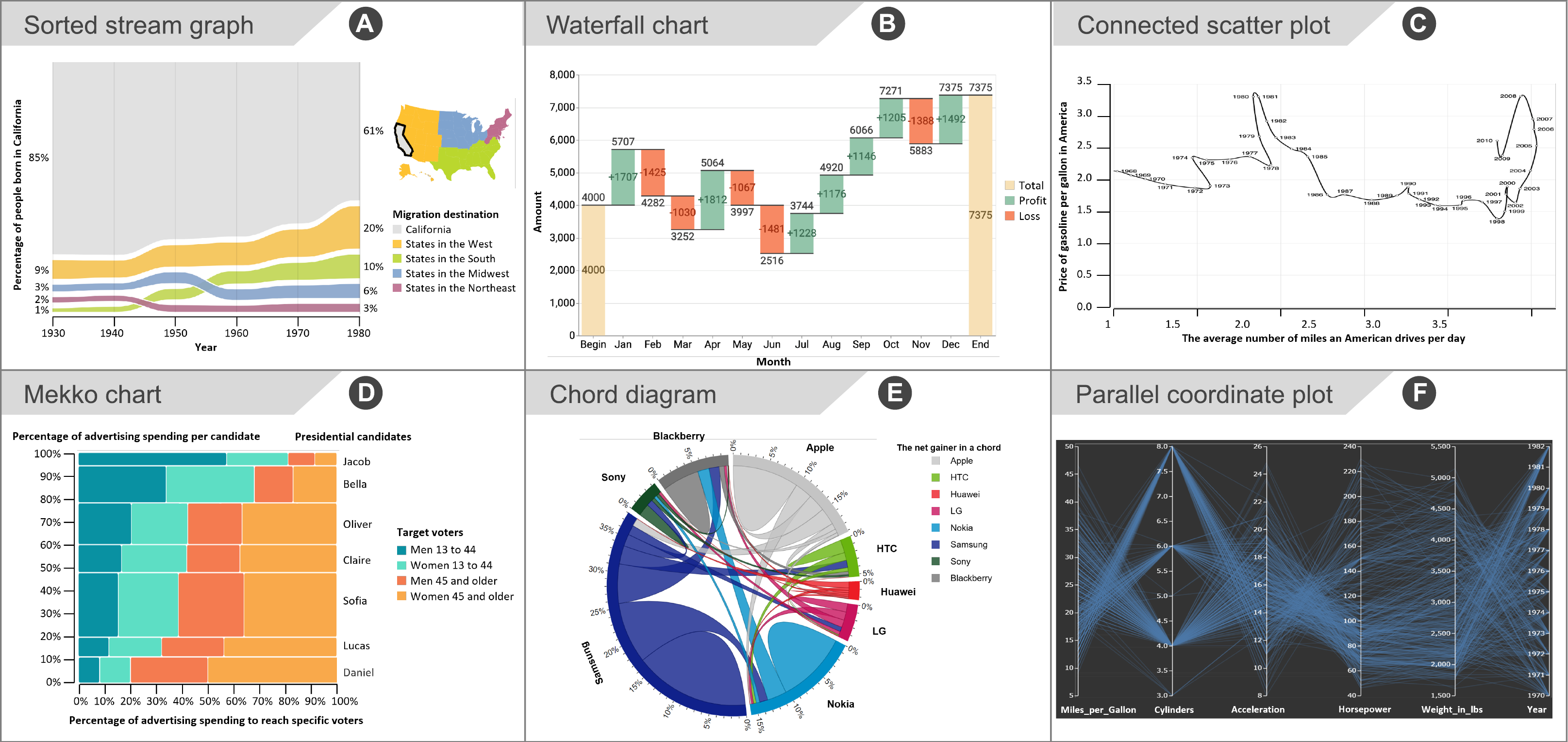}
  \caption{This figure shows the visualizations we used in our studies. A-D were used in Study 1 and C-F were used in Study 2 and Study 3.}
  \label{fig:visualizations}
\end{figure*}

We review approaches to facilitating understanding of visualizations and their applicability to oral presentations with visualizations.

\subsection{Techniques to Help Comprehend Visualizations}
The visualization community has attached importance to efficiently and effectively helping people understand visualizations to increase visualization accessibility for the general public.
Much research falls on evaluating people's ability to comprehend a visualization (\ie visualization literacy~\cite{borner2019data, lee2016vlat}) and improving visualization comprehension via training and education (\eg~\cite{alper2017visualization,chevalier2018observations}). For example, providing an explanation of a visualization followed by exercise questions has been shown to increase readers' comprehension~\cite{tanahashi2016study}, along with interactive tutorials~\cite{kwon2016comparative} and educational animations~\cite{ruchikachorn2015learning}. 
More recently, researchers have proposed cheat sheets that integrate graphical explanations such as data comics and textual annotations to help people understand visualization techniques~\cite{wang2020cheat}.
However, these works focus on educating visualization consumers and improving visual literacy in the long run.
Other existing works have identified techniques such as incorporating introductory text~\cite{fox2018read}, annotations (visual guides that walk people through the key points of a visualization), and summaries to help readers more efficiently extract takeaways from visualizations ~\cite{segel2010narrative}. 
Yet, they do not provide actionable guidelines to help presenters prepare the audience to better understand visualizations in live situations.

Another line of research that aims to improve visualization comprehension focuses on generating design guidelines for the visualizations themselves.
People can more accurately and quickly complete certain tasks with some visualization types than others. For example, people can more effectively extract trends from line charts and spot outliers from scatterplots~\cite{saket2018task}. 
Besides, adding a progress bar helps comprehend a data-GIF~\cite{shu2020makes}.
Additionally, highlighting key features of a visualization can facilitate understanding of articles with visualizations~\cite{zhi2019linking}. 
In contrast, existing research has found no significant improvement of comprehension when highlighting related parts of a visualization with visual cues in \textit{oral} presentations~\cite{kong2019understanding,kong2017internal}.
Overall, although these studies have generated concrete visualization design guidelines to help people create better visualizations, in a presentation setting, if the audience does not understand or misunderstands what is being presented due to the lack of an effective introduction, they could still struggle to extract insights from the visualizations. 

%%%%%%%%%%%%%%%%%%%%%%%%%%%%%%%%%%%%%%%%%%%%%%%%%%%%%%%%%%%%%%%%%%%%%%%%%%%%%%%%%%%%%%%%%%%%%
\subsection{Verbal Introductions}
The purpose of an introduction to visualization is to fill a knowledge gap between the designer's intention and the audience's interpretation. It increases clarity and improves comprehension~\cite{stoiber2019visualization}. 
For example, an effective introduction would convey the meaning of the visual encodings to the audience, since what each visual mark represents is at the discretion of the visualization designer. 
Referencing research from education, including concrete examples ~\cite{shimoda1993effects,sadoski2000engaging,sadoski1993causal,sadoski2001resolving,lefevre1986written} and topic sentences ~\cite{lorch1985topic,lorch1996effects,schwarz1981text} in introductions can facilitate comprehension of text-based materials. 
In this paper, we extend these previous findings in education to inform the visualization community by exploring whether these same techniques could also be applied to creating effective verbal introductions for visualizations. 

More recently, visualization researchers conducted case studies examining how to write effective introductory descriptions for triangular charts~\cite{fox2018read}. 
These charts used three sides of a triangle as three axes to present multidimensional data. 
The researchers evaluated two types of descriptive introductory text: the ``what-text'', which described the meaning of each component of a visualization (e.g., ``A point is an interval of time''), and the ``how-text'', which provided rules for extracting data from a visualization (e.g.,``Start-time: follow the left-most diagonal grid line to the intersection with the x-axis''). 
However, they found no significant differences between the two introduction types in facilitating understanding. 
To further investigate what constituted an effective introductory description, we generated several different introduction types, inspired by works in education research and observations from crowdsourced introductions, and examined their effectiveness in increasing visualization comprehension. 

%% file: section/3-pilot-study.tex
\section{Study 1: Crowdsourced Introductions}
\label{sec:study1}

\begin{table*}[t]
  \centering
  \includegraphics[width=\linewidth]{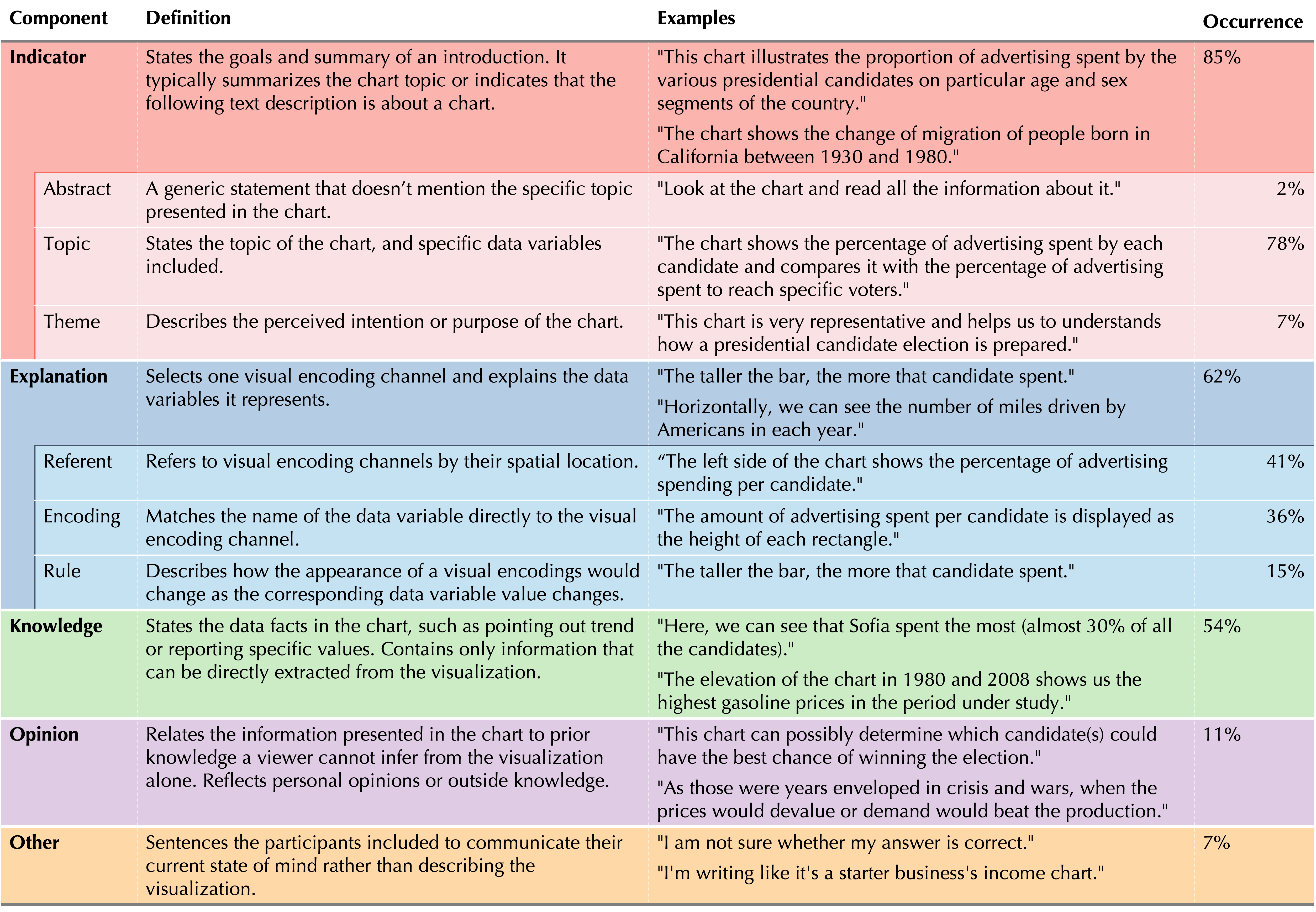}
  \caption{The categorization system, which identifies five components of an introduction (\ie \textit{Indicator}, \textit{Explanation}, \textit{Knowledge}, \textit{Opinion} and \textit{Other}) extracted from our crowdsourced data. Indicators and Explanations are further categorized into subcategories. The last column shows the percentage of introductions that includes each component or subcategory of a component.} 
  \label{tab:code_system}
\end{table*}

\rone{We started with understanding how members from the general public introduced a visualization via crowdsourcing, for two reasons. First, we were inspired by the use of ``wisdom of crowds'' that suggested combining opinions from crowds could drive better decisions than relying on that of individuals or even experts~\cite{surowiecki2005wisdom}. Second, this approach allowed us to survey introductions written by people with different levels of visual literacy to cover a range of possible approaches~\cite{hinds2001bothered,isaacs1987references}. 
Existing work had demonstrated that experts might not always have a good intuition for what non-experts saw in a visualization~\cite{xiong2021visual}, so exploring how people with varying levels of visual literacy introduced visualizations could potentially help us identify useful approaches in communication. 
We discussed the difference between introductions from participants with high and low visualization literacy in Section~\ref{discussionofpilot}. These crowdsourced introductions served as a corpus to map out a data-driven, inclusive design space of visualization introductions. From this design space, we generated several different introduction techniques, of which we examined the effectiveness in Study 2 and Study 3.}

\subsection{Study Design}
In the experiment, we showed participants four visualizations (a Mekko chart, a connected scatter plot, a sorted stream graph, and a waterfall chart). We asked them to compose an introduction to describe the visualization. We applied a coding scheme to analyze the introductions written by the participants. Based on this coding scheme, we created a categorization system that broke an introductory description down into meaningful units based on their content, structural purpose, relationship with other meaningful units, and the sequence in which they appeared.

\textbf{Visualizations.}
We selected visualizations in a moderate level of complexity to encourage participants to put more effort into their introductions. Too commonplace visualizations (e.g., bar charts) might not elicit interesting introduction strategies from participants, and participants might fail to comprehend extremely complex visualizations, producing incorrect introductions. 
We referenced visualizations from various sources, including news websites (e.g., New York Times~\cite{nyt} and Financial Times~\cite{fs}), galleries of visualization tools (e.g., D3~\cite{d3} and Vega~\cite{vega}), and visualization community websites (e.g., information is beautiful~\cite{infobeauty}).
We ended up selecting a Mekko chart~\cite{mekko}, a connected scatter plot~\cite{cscatter}, a sorted stream graph~\cite{stream}, and a waterfall chart~\cite{waterfall}, as shown in~\autoref{fig:visualizations}. 
We made sure the topics of these visualizations did not require experts' domain knowledge to accommodate participants with varying backgrounds. 
% via some pilot studies. Details can be found in the supplementary materials. 
For each visualization, we removed titles and annotations to prevent participants from copying or relying on them in their introductions. We excluded interactive components of the visualizations to focus on static visualizations. Finally, we modified the data labels in the visualizations to reduce the possibility that participants would write introductions based on their existing beliefs or biases due to controversies. 
The participants were told that the visualizations were modified and that they should not apply the information presented to real life.

\textbf{Procedure.}
Participants were randomly assigned to view one of the four visualizations. They were first asked to write down three findings or conclusions that they observed from the visualization. 
This task was for prompting the participants to read carefully.
Participants then wrote an introduction for the visualization. They were instructed to imagine that they made the visualization to convey some data in a presentation, and they had to write a script on what they would say to introduce the visualization so that the audience could understand how to read it. They were explicitly informed that the target audience had various backgrounds and visualization literacy levels.
\rthree{The study finished with a few demographic questions and a self-evaluated visualization literacy test developed by Garcia-Retamero et al.~\cite{garcia2016measuring}. Though the test was a self-assessment, Garcia-Retamero et al. had demonstrated that the subjective test could predict the literacy level as well as the objective test developed by Galesic and Garcia-Retamero~\cite{galesic2011graph}.}
The study took approximately 15 to 20 minutes, and participants were compensated at a rate of \pounds 7.50 per hour.

\subsection{Participants}
We recruited participants from the crowdsourcing platform Prolific~\cite{prolificwebsite}. We filtered for participants who were fluent in English and excluded everyone who demonstrated low effort or wrote clearly incorrect information. We ended up with 110 valid responses from participants ($60\%$ males, $37.2\%$ females, $0.9\%$ non-binary/third gender, $1.8\%$ prefer not to say, aged between 18 and 65, $\mu = 25.28 , \sigma = 8.57$), with 30 responses for the connected scatter plot and waterfall chart, and 25 responses for the Mekko chart and sorted stream graph. Their mean self-evaluated visualization literacy score was 20.28 out of 30 ($\sigma = 5.09$).

\subsection{Categorization System}
\label{code_system}
We qualitatively coded the collected introduction scripts, identified the distinct components, and categorized some strategies participants used in their introductions.
We decomposed each introduction script into meaningful units for analysis by using the end of lines and conjunctions (e.g., ``for example'', ``but'', ``and'') as signals of separation.
To classify those meaningful units, we iteratively reviewed and revised the coding scheme through several discussions and brainstorming with the card sorting and affinity diagram methods. 
After the coding scheme was determined, the first and third authors coded the introductions. 
We also recruited three external researchers to help with the coding, and the results were compared, with any differences resolved via discussions.

\begin{figure}[t]
  \centering
  \includegraphics[width=\linewidth]{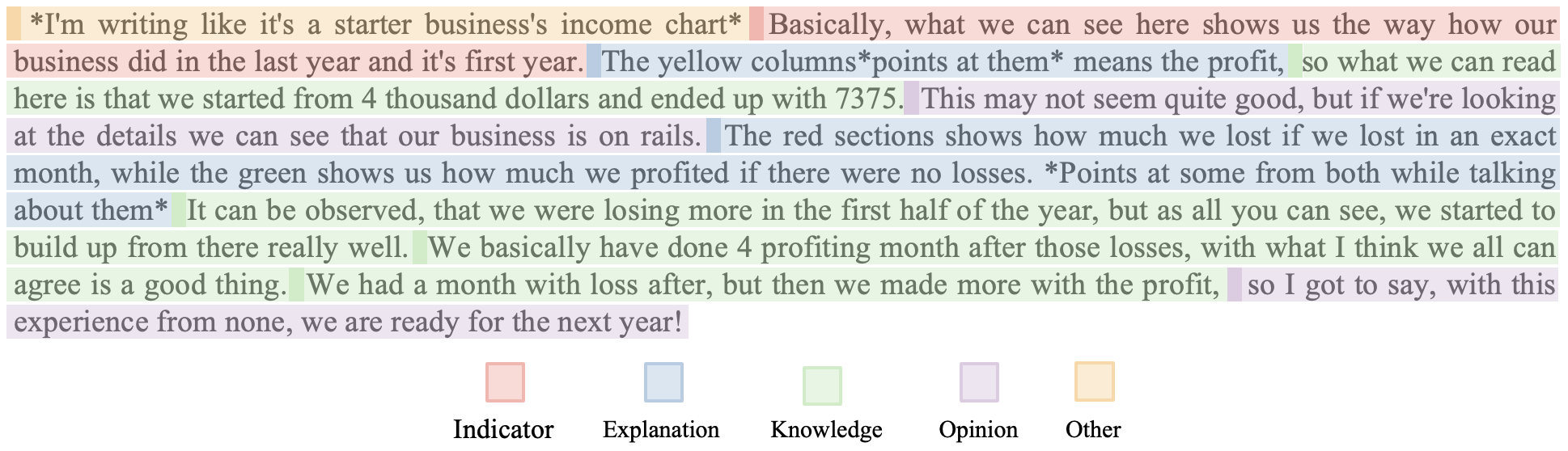}
  \caption{An example showing how the text categorization system works for an introduction of the waterfall chart. In this example, the type of the indicator component is topic, and the type of the explanation components is encoding.}
  \label{fig:code_sample}
\end{figure}

Our final categorization system classifies the meaningful units of introductions into five components: \textit{Indicator}, \textit{Explanation}, \textit{Knowledge}, \textit{Opinion}, and \textit{Other}, as shown in~\autoref{tab:code_system}.
~\autoref{fig:code_sample} shows an example of how we break down an introduction based on the categorization system.
To account for different narrative techniques participants use to express each component, we further categorize the indicator and explanation components into subcategories. 
\rone{For example, the three types of explanation (\textit{Referent}, \textit{Encoding}, \textit{Pattern}) express the same information of how to interpret the visual encoding channels, but in different ways. Referent explains visual encoding channels by referring to axis, text labels, and legends, Encoding identifies the visualized data variables and matches them to encoding channels, while Pattern identifies the encoding channels and then describes how they change as the corresponding data values change.}

\subsection{Results}

We analyzed both the content and structures of the introductions participants wrote. 
The insights we gained informed the design of our subsequent studies, where we generated and evaluated different types of introductions based on the categorization system. 
We excluded the opinion and other components from the following studies because we wanted to compose neutral introductions for a general audience. 
The domain-specific knowledge typically included in the opinion component was out of the scope of the present study, and the other component did not contribute to the effectiveness of an introduction.

\subsubsection{Components of Introductions} 
The three major components of introductions are: \textit{Indicator}, which states the goal or summarizes the introduction; \textit{Explanation}, which illustrates how visual encoding channels map onto data values; and \textit{Knowledge}, which describes facts and insights from the visualization. 
Some components are similar to what education literature has found that could enhance the understanding and retention of text. 
Specifically, the indicator component at the beginning of an introduction can serve as the topic sentence~\cite{lorch1985topic,lorch1996effects,schwarz1981text}, and the knowledge component is similar in effects to concrete examples~\cite{shimoda1993effects,sadoski2000engaging,sadoski1993causal,sadoski2001resolving,lefevre1986written}. 
On the other hand, the subcategories of indicator and explanation components are unique to the context of introducing visualizations.
% We report summary statistics on the frequency of each component and its subcategories included in introductions. This helps us identify types of content that most participants believed could enhance the effectiveness of their introductions.

The indicator component appeared the most frequently: 85\% of the participants wrote at least one indicator in their introductions before elaboration. Two participants wrote indicators in both the beginning and the end.
\textit{Topic} was the dominating indicator type. It stated the topic and data variables presented in the chart and appeared in 78\% of the introductions, followed by \textit{Theme} (7\%) and \textit{Abstract} (2\%). For explanation components, 62\% of participants wrote at least one explanation, while 48\% wrote two or more, totaling 172 instances of explanations observed in our dataset.
Participants wrote referent and encoding explanation types slightly more often than pattern explanation type, and they appeared in 41\%, 36\%, and 15\% of the introductions, respectively. 
We found that 25\% of participants used more than one explanation type in their introductions, suggesting that participants intuitively used multiple strategies to write an effective introduction. We documented 134 total instances of knowledge components in our dataset. 54\% of participants employed at least one knowledge component, and 28\% of them presented two or more.
% \autoref{tab:code_system} summarizes our categorization of the components and their frequencies in crowdsourced visualization introductions.

\subsubsection{Introduction Structure}
\label{study1structure}

\begin{figure}
    \centering
    \includegraphics[width=\linewidth]{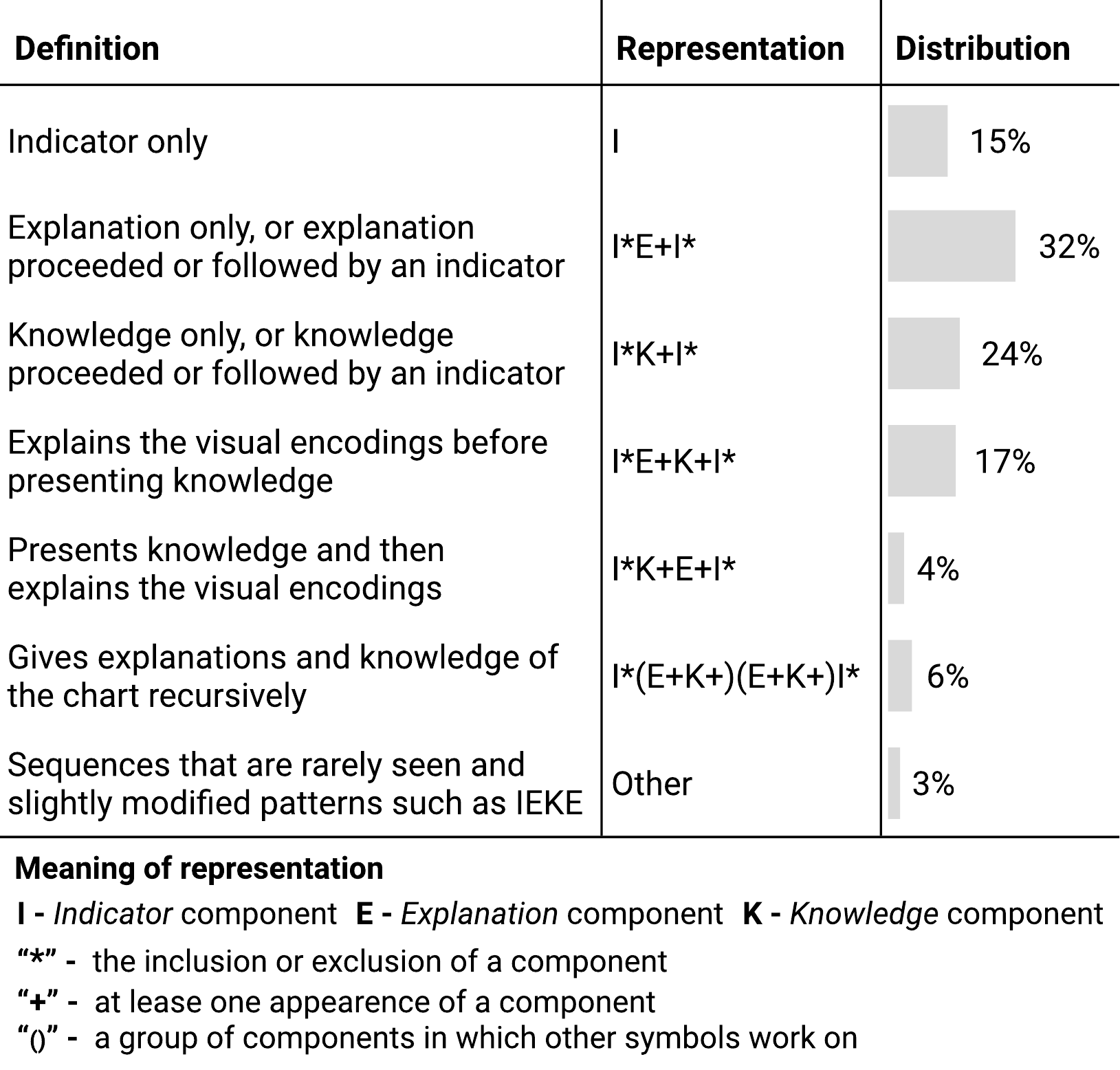}
    \caption{The frequencies of introduction structures in Study 1.
    }
    \label{fig:structure}
\end{figure}

We examined the order each component was presented in the visualization introductions. 
This analysis helped us identify the common frequencies and the sequences each component appeared in an introduction. 
For simplicity, we denoted \textit{\textbf{I}ndicator} as \textbf{I}, \textit{\textbf{E}xplanation} as \textbf{E}, and \textit{\textbf{K}nowledge} as \textbf{K}. An introduction structure was coded as a sequence of the characters. For example, if an introduction provided one explanation component, followed by a knowledge component, its structure would be equivalent to \textit{EK}. To improve the interpretability of our analysis, and to avoid a combinatorial explosion in the number of potential sequences, we merged consecutive components.  For example, if an introduction provided two explanation components, followed by a knowledge component, we would count the introduction structure as (explanation + knowledge), instead of (explanation + explanation + knowledge). 

Inspired by the work of Amini et al.~\cite{amini2015understanding}, we followed rules based on the regular expressions to represent the introduction structure. Specifically, we used a plus sign (+) to symbolize that a component might be repeated. So in the above examples, (explanation + explanation + knowledge) and (explanation + knowledge) could be represented by \textit{E+K}. 
Additionally, because 85\% of the participants included an indicator in their introductions, and all but one placed it at the beginning or the end, we combined the instances when an indicator was and was not included.
We used an asterisk (*) to reflect the inclusion and exclusion of indicators. For example, the \textit{I*E+I*} introduction type consisted of the following variations: \textit{IE+}, \textit{IE+I}, \textit{E+I}, and \textit{E+}.  

We discovered six unique structures. The distribution of their frequencies is shown in \autoref{fig:structure}.
The two most common structures were \textit{I*E+I*} (32\%) and \textit{I*K+I*} (24\%) as the majority of the participants only included either the explanation or the knowledge component in their introductions.
We also observed a substantial amount of introductions (27\%) that consisted of both explanation and knowledge components. 
Interestingly, most participants intuitively presented explanation components first, then knowledge components (\ie \textit{I*E+K+I*} (17\%)), rather than the other way around (\ie \textit{I*K+E+I*} (4\%)).
There were also instances where participants interwove the explanation and knowledge components in their introductions (\ie \textit{I*(E+K+)(E+K+)+I*} (6\%)). For example, one could write an explanation of one visual encoding channel (e.g., color), immediately followed by a knowledge component that served as an example to the encoding channel explanation (e.g., describing the trend in data for a specific color group). To account for this structure, we used a pair of parentheses () to indicate \textit{related} explanation and knowledge components. 
The introduction script in \autoref{fig:code_sample} is an example of this structure, which includes two pairs of related explanation and knowledge components.

\subsection{Discussion}
\label{discussionofpilot}
\rthree{All of the three explanation and component types were seen in the introductions for the four visualizations. In terms of introduction structure, the most frequent structures \textit{I*E+I*}, \textit{I*K+I*}, and \textit{I*E+K+I*} were seen in the introductions in all four charts. We found that \textit{I*E+I*} was the most or second most frequent structure overall, suggesting that the explanation component was the most important element of an introduction. The major difference was in the use of \textit{I*K+I*} and \textit{I*E+K+I*}. Specifically, the introductions for the stored stream graph and waterfall chart used \textit{I*K+I*} more frequently than \textit{I*E+K+I*}. However, the introductions for the connected scatter plot and the Mekko chart used \textit{I*E+K+I*} more often. It implied that participants considered the Mekko chart and the connected scatter plot to be more difficult to understand, and made a conscious effort to leverage the explanation components to increase the effectiveness of their introductions.}

To compare introductions written by participants with different visualization literacy levels, we used the Pearson correlation coefficient to investigate the relationship between participants' literacy scores and the proportion of different components in their introductions.
We found a weak positive relationship ($Pearson's \ r = 0.27$) between literacy scores and the proportion of explanation components in their introductions. The participants with higher literacy scores tended to include more explanation components in their introductions.
We also found a weak negative relationship ($Pearson's \ r = -0.22$) between literacy scores and the proportion of knowledge components, such that participants with higher literacy scores tended to include fewer knowledge components. 
This might suggest that participants with higher visualization literacy levels preferred writing introductions with the explanation component, which was more abstract, more so than the knowledge component, which was more concrete. 
This echoed prior research in communication, which found that experts tended to use more abstract statements in communication compared with non-experts~\cite{hinds2001bothered}. 
On the other hand, the indicator component was almost equally used by participants with varying levels of visualization literacy ($Pearson's \ r = -0.07$).

Further, we found some participants gave incorrect answers when they were asked to write three findings or conclusions they observed from the charts (average number of correct answers: Mekko chart: 2.24 out of 3, connected scatter plot: 2.57, waterfall chart: 2.63, and sorted stream graph: 2.68). This suggested that not all participants were able to fully understand the visualizations without an introduction. We examined the introductions they wrote nonetheless, to cover as wide of a range of strategies as possible to create our strategy categorization system. This did not impact the validity of our system because it focused on documenting strategies, which had more to do with \textit{how} components were introduced than \textit{what} components were introduced, making this system a valid one independent of the crowd workers' knowledge of the chart.

Next, based on these crowdsourced results, we generated new introductions for visualizations to compare their effectiveness in facilitating comprehension. Because of many approaches available to create an introduction, to avoid a combinatorial explosion in the number of comparisons, we focused on identifying the most effective \textit{explanation component (E)} in Study 2. Then, in Study 3, we examined the effectiveness of the inclusion of an indicator and the inclusion of knowledge components. We also identified the most effective sequence of components in an introduction. 
% This investigation can help us generate concrete guidelines to help visualization presenters compose an effective introduction for their visualization. 
% This investigation will supplement existing findings from education to see if including topic sentences (indicator components)~\cite{lorch1985topic,lorch1996effects,schwarz1981text} and giving concrete examples (knowledge components)~\cite{sadoski2000engaging,sadoski1993causal,sadoski2001resolving}, which have been shown to facilitate understanding in reading text-based materials, can also help people better understand visualizations. 

%% file: section/4-study1.tex
\section{Study 2: Evaluate Explanation Types}
\label{sec:study2}
In this study, we compared the three approaches to \textbf{the explanation component} to examine which was the most effective. 
% at facilitating people's understanding of visualization.

\subsection{Study Design and Procedure}
Participants were randomly assigned to view one of four visualizations while listening to an audio clip introducing the chart using one of the three explanation types (Referent, Encoding, Pattern). The introduction was given through audio to mimic the oral presentation scenarios~\cite{kong2019understanding,kong2017internal}. After the audio ended, the participants completed a comprehension test with eight questions about the chart they saw. The questions were presented one at a time in random order. 
In the end, participants completed a subjective visualization literacy assessment~\cite{garcia2016measuring} and answered some demographic questions. 
\rone{Participants also reported whether they had seen the chart in the experiment before and how often they encountered such chart types in their life.}
The study took approximately 10 minutes, and participants were compensated at a rate of \pounds 7.50 per hour. 

We selected a chord diagram~\cite{chorddiagram} and a parallel coordinated plot (PCP)~\cite{parallelcoordinate}, together with the Mekko chart and connected scatter plot from Study 1 to conduct experiments. 
We selected these visualizations based on their visual complexity and use of different encoding channels. We tested multiple visualization types in Study 1 and found that participants tended to struggle more understanding Mekko charts and connected scatter plots (average number of correct answers from Study 1: Mekko chart: 2.24 out of 3, connected scatter plot: 2.57, waterfall chart: 2.63, and sorted stream graph: 2.68). We also used the PCP and chord diagram because they were less familiar to the general public and were often misunderstood by readers~\cite{lee2015people}. Previous studies also found that the connected scatter plot and PCP could cause misinterpretations~\cite{haroz2015connected, kwon2016comparative}. 
Together, these visualizations were great test beds because introductions would likely have a larger effect helping people understand them, and any improvements in comprehension would likely be more saliently detected. Additionally, these charts visualized diverse data types (i.e., hierarchical, time series, multivariate, and network data) with different encoding rules, so this enhanced the generalizability of our study results.

We also piloted our comprehension questions in this study with 120 participants to avoid floor and ceiling effects. Participants also provided feedback on the experiment instruction and the audio. Afterward, we added a volume test in the survey for participants to adjust the volume of their computers before starting the experiment. 
\subsubsection{Audio Explanations}
The audio explanations the participants listened to were designed to mimic the real-world scenario of listening to a verbal presentation.
\rthree{The audio scripts for the three explanations types were composed to have similar complexity and length. We iteratively revised our scripts based on feedback from external researchers who were unfamiliar with our work, until all three were perceived to be clear and similarly complex.}
\rtwo{The average duration of the audio of the three explanation types were 39 seconds (Referent), 38 seconds (Encoding), and 47 seconds (Pattern). Note that these explanations were \textit{based on} the structures we observed from the crowdsourced introductions (see Table \ref{tab:code_system}), rather than a direct replication of what those participants wrote.}
Note that neither indicator nor knowledge components were included so we could isolate the effect of the different types of explanation component. 

\subsubsection{Participants}
We recruited 360 participants from Prolific~\cite{prolificwebsite}, with 30 participants in each condition. We filtered participants who jumped to the comprehension test section before the audio ended, finished the comprehension test too fast (completion time $Z-score < 3$), or gave wrong answers clearly indicating they did not read the questions (e.g., giving ``Jacob'' as the answer for each question about the Mekko chart). There were 347 valid responses, with ages ranging from 18 to 75 ($\mu = 34.45, \sigma = 13.74$, $59.65\%$ females, $39.48\%$ males, $0.86\%$ non-binary/third gender). Their average visualization literacy score was 18.9 out of 30 ($\sigma = 5.33$).

\subsubsection{Comprehension Test}
We used the comprehension test scores of participants as the metric of the effectiveness of the explanations.
The comprehension test consisted of eight questions to cover a wide range of task types within a reasonable time.
The questions were based on the VLAT, the visualization literacy assessment framework developed by Lee et al.~\cite{lee2016vlat}, which suggested designing questions based on different visualization tasks such as \textit{Retrieve Value}, \textit{Find Extrema}, and \textit{Make Comparisons}. The design of questions for the connected scatter plot and PCP also referenced common misinterpretations identified in the work by Haroz et al.~\cite{haroz2015connected} and the work by Kwon et al.~\cite{kwon2016comparative}, respectively.
In developing the questions, we noticed that the \textit{Make Comparison} questions in VLAT were True or False questions, yielding a 50\% random guessing accuracy. Thus, we modified the \textit{Make Comparison} questions to a short answer format, which required making comparisons of multiple data value pairs instead. 

We also balanced the difficulty levels of our comprehension questions following guidelines from VLAT~\cite{lee2016vlat}, which ended up with two easy, three medium, and three difficult questions. Taking questions for the Mekko chart as an example, the easy questions asked participants to retrieve specific values from the chart, such as ``What is the percentage of Claire's advertising spending?'' The medium questions asked participants to make comparisons before retrieving specific values, such as ``Which group of voters does Jacob spend the most on?'' and ``In terms of overall advertising spending, who spent the most on men voters 13 to 44?'' The difficult questions required participants to further filter for specific criteria before comparing and retrieving values, such as ``In terms of overall advertising spending, who spent most on men voters?''

\subsection{Results}

\begin{figure}[t]
  \includegraphics[width=\linewidth]{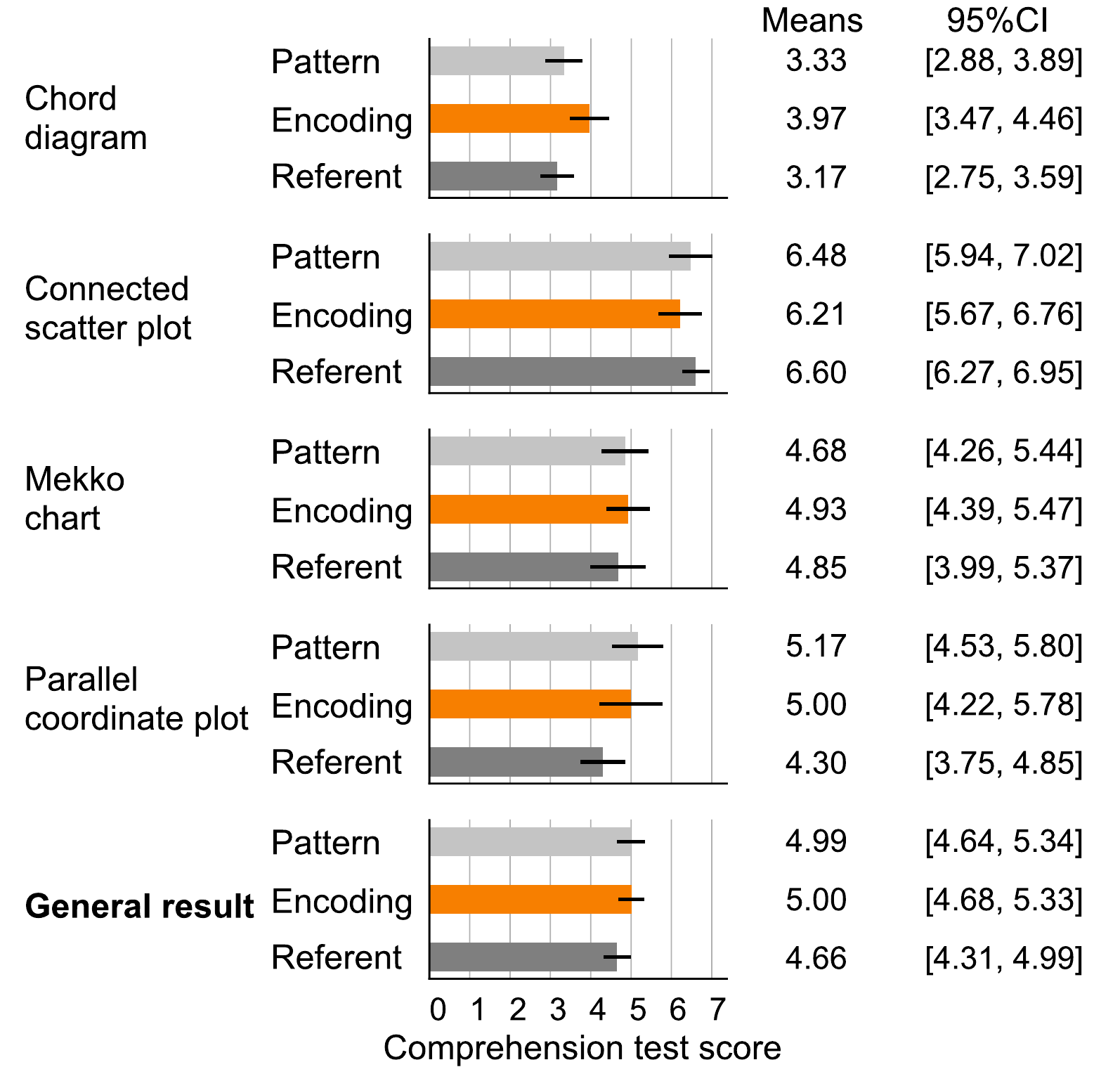}
  \caption{The chart shows the average comprehension test score for each visualization and explanation type. There was no one significantly best explanation type that yielded the best results across all chart types. 
}
  \label{fig:exp1result}
\end{figure}

\subsubsection{Effects of Explanation Type}
We used a two-way ANOVA with chart type and explanation type as independent variables, and participants' performance on the comprehension test as the dependent variable. 
We found no significant effects of explanation type on the comprehension test scores ($F = 1.73,\ p = 0.18,\ {\eta^2}_{partial} = 0.01$), but a significant effect of chart type ($F = 52.46,\ p < 0.001,\ {\eta^2}_{partial} = 0.32$). The average comprehension test scores in each explanation type are shown in~\autoref{fig:exp1result}. Post-hoc comparison using Tukey's HSD test suggested that participants performed worst in the comprehension questions with the chord diagram (p-value all less than $0.05$). Among the connected scatter plot, Mekko chart, and PCP, participants performed better with the connected scatter plot than the Mekko chart and PCP (p-value all less than $0.05$). No difference were found between the Mekko chart and PCP. No interaction between chart type and explanation type was found ($F = 1.29,\ p = 0.26,\ {\eta^2}_{partial} = 0.02$). 
% The average comprehension test scores in each explanation type were: referent ($\mu = 4.66,\ CI_{95\%} = [4.31,4.99],\ \sigma = 1.89$), encoding ($\mu = 5.00,\ CI_{95\%} = [4.68,5.33],\ \sigma = 1.79$), and pattern ($\mu = 4.99,\ CI_{95\%} = [4.64,5.34],\ \sigma = 1.87$). Participants listening to the encoding and pattern type performed insignificantly better.

\rone{Participants were also asked to indicate whether they had seen the charts used in the experiment before by selecting from ``Never before'', ``Rarely'', and ``Sometimes''. All participants selected ``Never before''. We conducted a two-way ANOVA to examine the effect of the chart type and the frequency of seeing in participants' performance. We found no significant effects of the frequency of seeing the chart type on the comprehension test scores ($F = 0.12,\ p = 0.89,\ {\eta^2}_{partial} = 0.001$).}

\subsubsection{Effects of Visualization Literacy}
As a sanity check, we compared the visualization literacy scores between participants who viewed different explanation types via a one-way ANONA. We found no significant difference in the literacy scores between participants who viewed different explanation types ($p = 0.55,\ F = 0.60,\ {\eta^2}_{partial} = 0.003$). We conducted a linear regression analysis to examine the effects of visualization literacy scores, in addition to explanation type and chart type. 
The result showed that visualization literacy score was a statistically significant predictor of performance ($p < 0.001,\ coef = 0.083$), where higher visualization literacy was associated with better performance on the comprehension test.
In addition, we constructed a three-way ANOVA model with the literacy score, chart type, explanation type as independent variables. We observed similar effects of explanation type, chart type, and their interaction as previously reported in the two-way ANOVA.

\subsection{Discussion}
Our results suggested that the three explanation approaches did not seem to have a differentiable impact on comprehension. 
Additionally, when the instruction materials were similar enough, the participants' visual literacy levels became a critical determinant of how well they understood a visualization. Based on the coding system in Study 1, we could also investigate and compare the effectiveness of the different types of indicator components and different examples of knowledge components. However, based on the results from Study 2, we suspected the difference between each component type would, similar to the explanation component types, be minimal.
Next, we turned our attention to the overall structure of the introduction and investigated how the three major components interacted to influence the overall effectiveness of an introduction. 

%% file: section/5-study2.tex
\section{Study 3: Evaluate Structures}
\label{sec:study3}

\begin{figure*}[t]
  \includegraphics[width=\textwidth]{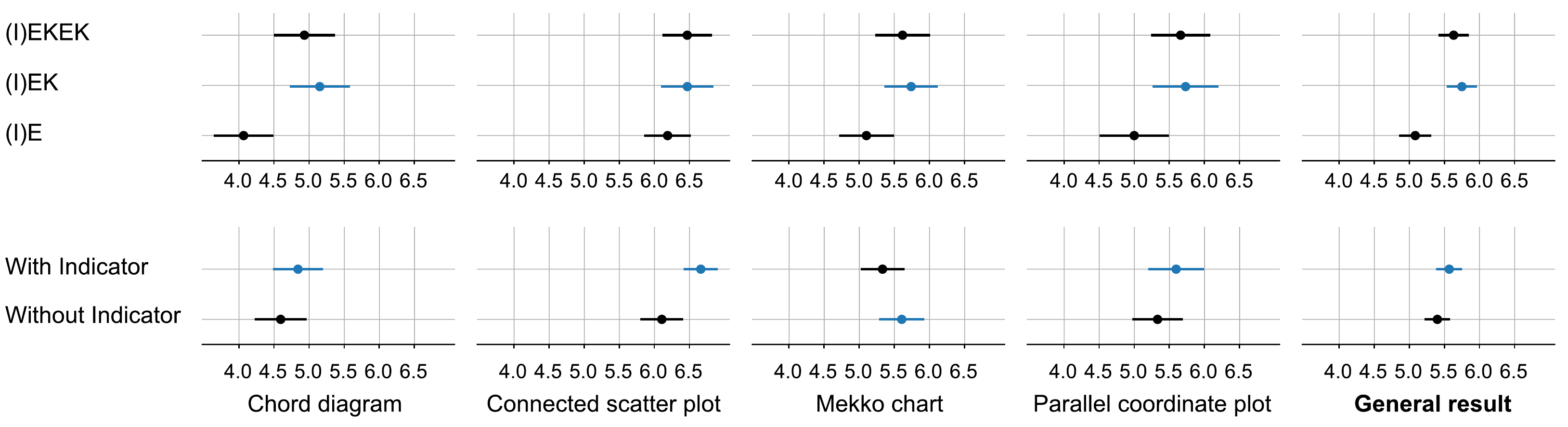}
  \caption{The chart shows the average comprehension test score and 95\% confidence intervals for each visualization and introduction type. The upper section shows the main effect of the body content (whether the knowledge component is included, and whether it is interwoven with the explanation component). The lower section compares introductions with and without the indicator component. Blue error bars represent introductions that yielded best results. Generally, \textbf{introduction type \textit{(I)EK} and \textit{(I)EKEK} are significantly better} than introduction type \textit{(I)E}. 
%   \aoyu{helpful to indicate which are significant} \done{}
}
  \label{fig:exp2result}
\end{figure*}

In this experiment, we investigated how the structure of an introduction affected the understanding of visualizations. We defined the structure of an introduction by the existence and order of a component---whether indicator, explanation, or knowledge was included or not, and the relative order of each component. However, permuting major components and subcategories of them to compare all possible structures was infeasible.
Instead, we limited our investigation scope by applying several restrictions and discussed not-examined conditions in Section~\ref{sec:limitation}. 
First, because indicator was the most popular component, we examined whether it improved the effectiveness of an introduction or not. 
We generated introductions that included an indicator component and introductions that did not. 
However, we limited our investigation to when the indicator component was added to the very beginning of the introduction, as it was the most popular approach. More specifically, among 94 participants whose introductions had at least one indicator component, 75 participants provided an indicator at the beginning of their introductions.     
We also examined the effect of including versus excluding the knowledge component. But we limited our investigation to the conditions where the knowledge component followed an explanation, or was interwoven with an explanation. We made these choices because they were some of the most common introduction structures we found in Study 1.

\subsection{Study Design and Procedure}
\label{exp2Design}
We created six introduction structure types: explanation only (\textit{E}), explanation followed by knowledge (\textit{EK}), explanation interwoven with knowledge (\textit{EKEK}), indicator plus explanation (\textit{IE}), indicator plus explanation followed by knowledge (\textit{IEK}), and indicator plus explanation interwoven with knowledge (\textit{IEKEK}). Notice that the last three introduction types were identical to the first three but with an indicator component added to the beginning. 
This setup allowed us to investigate the effect of including/excluding the indicator component by comparing the first three types to the last three types. 
We also investigated the effect of the knowledge component by comparing (1) the introduction types that excluded the knowledge components (\textit{E} and \textit{IE}), (2) the types that included knowledge but were not interwoven with explanation (\textit{EK} and \textit{IEK}), and (3) the types that interwove knowledge and explanation (\textit{EKEK} and \textit{IEKEK}). This between-subject study followed the same procedure as Study 2. Every participant was randomly assigned to listen to one of the six introduction types while viewing one of the four visualizations. The study took approximately 10 minutes, and participants were compensated at a rate of \pounds 7.50 per hour.

\subsubsection{Audio Introductions}
For the introductions where the knowledge component was included but not interwoven with the explanation (\textit{EK} and \textit{IEK}), the explanation component explained the meaning of all visual channels. Then the knowledge component gave examples afterward. For the introduction types where explanation and knowledge components were interwoven (\textit{EKEK} and \textit{IEKEK}), each explanation was immediately followed by a corresponding example. 
\rthree{For the same type of component in different introduction types, such as the knowledge components in introduction types \textit{EK}, \textit{IEK}, \textit{EKEK}, and \textit{IEKEK}, we made sure their contents were identical.}
\rtwo{The average duration of the audio of the introduction types were: 49 seconds (IE), 87 seconds (EK), 96 seconds (IEK), 91 seconds (EKEK), and 100 seconds (IEKEK).} For the explanation component of each chart, we chose the explanation type that yielded the best results (although insignificantly) in Study 2. Specifically, we chose the encoding explanation for the Mekko chart and chord diagram, the referent explanation for the connected scatter plot, and the pattern explanation for the PCP. For the indicator component, we chose the topic type, which was the most commonly observed in Study 1. For the knowledge component, we ensured that the data values related to the knowledge component were not the answers to any questions in the comprehension test to maintain fairness across all six introduction types.

% we selected the extreme data values on the chart as examples. For instance, a knowledge component describing the height of bars reads, ``For example, the fifth row representing Sofia is the tallest row among all rows. It means Sofia has spent more on advertising compared to other candidates." 

\subsubsection{Participants}
We recruited 720 participants from Prolific~\cite{prolificwebsite}, with 30 participants in each condition.
After applying the same exclusion criteria in Study 2, we ended up with 680 valid responses, with ages ranging from 18 to 86 ($\mu = 34.85, \sigma = 13.52$, $55.44\%$ females, $43.82\%$ males, $0.73\%$ non-binary/third gender). Their average visualization literacy score was 19.48 out of 30 ($\sigma = 5.04$).

\subsection{Results}
\subsubsection{Effects of Introduction Types}
\label{sec:study3result}
We examined the effects of introduction types by using a three-way ANOVA. The ANOVA had three factors: indicator inclusion, introduction body content, and chart type. Indicator inclusion had two levels: including the indicator (referring to introduction type \textit{IE}, \textit{IEK}, and \textit{IEKEK}) and not including the indicator (referring to introduction type \textit{E}, \textit{EK}, and \textit{EKEK}). Introduction body content had three levels: no knowledge (\textit{E} and \textit{IE}), separated knowledge (\textit{EK} and \textit{IEK}), and interwoven knowledge (\textit{EKEK} and \textit{IEKEK}). We also included the interaction term between every two factors. The overview of the experiment's results is shown in~\autoref{fig:exp2result}.

We found \textbf{no main effect of indicator inclusion} ($F = 2.61,\ p = 0.11,\ {\eta^2}_{partial} = 0.0039$), such that people did not score higher on the comprehension tests when the introduction included an indicator ($\mu = 5.57,\ CI_{95\%} = [5.39,5.75],\ \sigma = 1.70$) than when the introduction did not include an indicator ($\mu = 5.40,\ CI_{95\%} = [5.22,5.58],\ \sigma = 1.72$). 

There was \textbf{a main effect of introduction body content} ($F = 12.55,\ p < 0.001,\ {\eta^2}_{partial} = 0.037$). Post-hoc comparisons using Tukey's HSD test suggested that participants who listened to introductions with the knowledge components, regardless of whether the explanation and knowledge components were separated (\textit{EK} and \textit{IEK}) or interwoven (\textit{EKEK} and \textit{IEKEK}), performed significantly better than those who listened to introductions without the knowledge component (p-value all less than $0.001$). We found no significant difference between the separated knowledge and interwoven knowledge introduction types ($p_{adj} = 0.72$). The average comprehension test scores in each level were: no knowledge ($\mu = 5.09,\ CI_{95\%} = [4.86,5.31],\ \sigma = 1.78$), separated knowledge ($\mu = 5.75,\ CI_{95\%} = [5.53,5.97],\ \sigma = 1.64$ ), and interwoven knowledge ($\mu = 5.63,\ CI_{95\%} = [5.42,5.85],\ \sigma = 1.64$). 

There was also a main effect of chart type ($F = 31.27,\ p < 0.001,\ {\eta^2}_{partial} = 0.125$). Post-hoc comparisons found similar results to that in Study 2, where participants performed better with connected scatter plots, then with the Mekko chart and PCP, and the least well with the chord diagram. \rone{Again, no participant reported to had seen the charts in the experiment before. A two-way ANOVA examining the effect of the chart type and the frequency suggested there to be no significant effects of the frequency of seeing the chart on the comprehension test scores ($F = 0.24,\ p = 0.87,\ {\eta^2}_{partial} = 0.001$).} No interaction effect was found.

\subsubsection{Effects of Literacy Score}
As a sanity check, we compared visualization literacy scores between participants in different conditions via a two-way ANOVA. We checked the effect of indicator inclusion, introduction body content, and their interaction. We found no significant difference of literacy scores between participants who listened to introductions with different introduction body content ($p = 0.38,\ F = 0.97,\ {\eta^2}_{partial} = 0.002$). Also, there was no significant difference between participants who listened to introductions with and without an indicator ($p = 0.88,\ F = 0.02,\ {\eta^2}_{partial} < 0.001$). No interaction effect was found.

We constructed a linear regression model predicting performance on the comprehension test with the literacy score, chart type, indicator inclusion, introduction body content, and their interactions. We observed significant effects of visual literacy on comprehension test performance ($coef = 0.08,\ p < 0.001$). We found similar effects of indicator inclusion, introduction body content, chart type, and their interaction as reported in Section~\ref{sec:study3result}, suggesting that although visual literacy had predictive power in participant performance, the effect of indicator inclusion and knowledge component sequence still held.

%% file: section/6-discussion.tex
\section{Discussion}
Next, we discuss implications for designing verbal introductions of visualizations and future research in this area.

\subsection{Implication}
\rthree{\textbf{Get feedback from people with different levels of expertise.}}
In Study 1, we collected diverse approaches to introducing visualizations via crowdsourcing. 
Interestingly, we found that participants with high visual literacy tended to steer away from using concrete examples in their introductions, and instead relied on general, abstract rules to describe the visualizations. However, our experimental results showed that including the knowledge components (usually in the form of concrete examples) significantly improved participants' performance. 
This showed the potential benefits of combining opinions of people from diverse backgrounds to drive decisions on visualization-related designs.
It also indicated that when preparing presentations, presenters should remember that people with different levels of expertise have different communication preferences. Therefore, getting feedback on the presentation content from people with varying levels of expertise could help accommodate a broader audience.

\rthree{\textbf{Combine different approaches to giving an explanation of visual encoding channels.}}
In Study 2, we compared three types of explanation but found no significant difference between their effectiveness. Although not significantly, the encoding explanation yielded the best results for the Mekko chart and chord diagram, the referent explanation for the connected scatter plot, and the pattern explanation for PCP. 
This suggested that there was no one-size-fits-all solution to explain the visual encoding of visualizations to the audience. 
We would recommend that presenters consider combining different approaches in their introductions when a visual encoding channel is complex.
% Unexpectedly, we thought that the pattern explanation would be the most effective as it explicitly told how to interpret the visual patterns of visual encoding channels; however, such a more detailed explanation did not improve the comprehension. 
% It may indicate that what people could absorb from abstract explanation is limited, no matter how detailed the explanation is.

\rthree{\textbf{Give examples in introductions.}} 
Our Study 3 results showed that adding knowledge components to give examples after the explanation components was more effective than only giving abstract explanations. 
This supplemented the findings from research in education, where using concrete examples was shown to facilitate learning as examples evoked the mental imagery in learners to help decode abstract information in a meaningful way~\cite{sadoski2013imagery,bolkan2019examples,sadoski1993causal}, and increased learners' engagement~\cite{sadoski2000engaging,sadoski2001resolving}. 
While we found examples effective, it remained unclear what factors could further enhance examples, such as what kinds of data facts could enhance the imagery and memorability of an example.
For instance, for the Mekko chart, the majority used the rows representing data on ``Jacob'' and ``Daniel'' as examples. They happened to be the top and bottom rows in the chart, respectively. 
They could be selected as examples because they were easy to locate in the chart, or because they were the two extrema in the chart, and thus being very salient.
This suggested that there might exist rules or tendencies for people to select specific patterns in a visualization as the concrete examples in their introductions. 
To generate more specific guidelines on giving examples, future research could further examine these rules and tendencies.
\rtwo{Additionally, other visual features (e.g., salient visual patterns and textual information) might distract the audience's attention. Future work could investigate how examples would interact with these other visual features to impact comprehension.}

\rthree{\textbf{Do not rely on the indicator component to convey massages.}}
Previous studies suggested that the topic sentence helped readers grasp the main idea and structure of a text paragraph and further enhanced their retention and comprehension~\cite{lorch1985topic,lorch1996effects,schwarz1981text}.
However, in our experiments, when an indicator component was given at the beginning of introductions to summarize the topic of the chart, its effect in enhancing visualization comprehension was negligible. 
The contradiction might come from the difference between how people processed audio and textual information. 
Readers could repeatedly read a text and draw connections between the topic sentence and the body text to better understand the content. However, doing so when listening to an audio clip required the participants to hold the topic sentence in memory and process the rest of the introduction, which could be cognitively demanding. 
Future studies could investigate other ways to provide topic sentences in a visualization introduction, such as utilizing titles or annotations. 

\rthree{\textbf{Prevent misinterpretations from different aspects.}}
Through a close look at participants' introductions in Study 1, we found that their misinterpretations came from different sources.
First, participants' existing knowledge of conventional design could influence how they understand unconventional ones, echoing previous research findings~\cite{fox2018read}.  
For example, as shown in~\autoref{fig:visualizations}B, the waterfall chart showed a company's cumulative net income in each month through a rectangle. A red rectangle represented \textit{a loss of income} in that month. The highest point of a rectangle represented the total net income at the beginning of the month. Its length represented the amount of loss that month. The lowest position represented the total net income at the end of the month.  
Several participants in our study misread the position of the top of the red rectangles as the total net income in that month, as if they were reading a regular bar chart.
This suggested that when unconventional charts shared a visually similar design with familiar ones, people might make wrong assumptions about the unconventional charts based on what they already knew. 
Presenters could design introductions to mitigate these mistakes by explicitly mentioning the differences between a novel chart and conventional charts that share similar design elements. 
Additionally, participants could be careless when they were reading visualizations and misunderstood nuances. 
For example, one participant stated that the sorted stream graph was about \textit{``a percentage of people born in California and some other states''} while the text labels indicated that the data was about the percentage of people \textit{born in California} that stayed in California or immigrated to other states. 
When designing introductions, presenters could re-emphasize key ideas of the chart. Future research could identify and document possible sources of chart misunderstanding to mitigate their effects.

%%%%%%%%%%%%%%%%%%%%%%%%%%%%%%%%%%%%%%%%%%%%%%%%%%%%
\subsection{Limitation and Future Directions}
\label{sec:limitation}
\textbf{Investigate other types of verbal introductions.} Our paper only investigated a subset of introduction types from our classification system. 
% We only compared the effectiveness of the three explanation component subcategories on comprehension in Study 1, and we only included the best performing explanation type in our introduction type evaluation in Study 2. 
Future work could compare the effectiveness of three indicator subcategories, other introduction structures, as well as the interaction effect between the included subcategory of a component and the introduction structure. For instance, one could compare the effectiveness of introductions having the same knowledge components but different explanation types to see whether certain explanation types would enhance the effect of adding knowledge components. 
\rtwo{Moreover, we did not include contextual information such as opinions and outside knowledge in our introductions. Future work could additionally incorporate these contextual information in introductions and investigate their impacts on the effectiveness of an introduction.}
Finally, for the current study, we did not consider other, less intuitive approaches to introduce a visualization, such as using colloquial language to tell an engaging narrative~\cite{kerby2018fusion}.
A recent study found that incorporating narrative introductions to a piece of text made the content more memorable for readers~\cite{mensink2021different}. Future research could investigate those other types of introductions and their effects on visualization comprehension.

\textbf{Reduce the effect of visualization literacy levels.}
Our experimental results showed that the visualization literacy levels accounted for some differences between participants' performance on comprehension tests, above and beyond the effect of introduction types. The importance of visualization literacy called for more research on promoting visualization education for the general public to facilitate democratizing data visualization. 
Future research could also study customizing the visualization and reading tutorials based on the audience's visualization literacy levels.

\textbf{Investigate other visualizations and dependent measures.} 
Our experiments used four charts that we identified as having the desired complexity for our investigation. Future research could explore how different introduction types would affect comprehension across a broader range of visualization designs.
Further, the present studies accessed the effectiveness of visualization introductions only via comprehension questions adapted from VLAT~\cite{lee2016vlat}. 
\rthree{However, some questions might be easier to answer with certain introduction types than others. For instance, a referent explanation that explained an axis might facilitate answering questions related to retrieving values. A pattern explanation might facilitate answering questions related to comparison tasks, since it explained how a visual encoding channel would change as the corresponding data values changed. 
% There could be other metrics to evaluate the effectiveness of an introduction. 
Future work could investigate the effectiveness of introduction types with other tasks, such as content recall~\cite{kong2019understanding, ajani2021declutter} or audience engagement~\cite{boy2015storytelling}.}

\textbf{Study more complex scenarios.}
\rtwo{In our experiment, participants saw a single static visualization accompanied by audio introductions. However, oral presentations could be more complex in real-world scenarios. 
For example, presenters might break down components of a complex visualization and reveal each component progressively through animation, present multiple related visualizations, or use deictic gestures to guide the audience's attention. Future research should explore the effect of visualization introductions in these more complex scenarios.}

\textbf{Use approaches beyond crowdsourcing.}
There were also some limitations of the crowdsourcing approach. 
First, we could not ask participants for their motivations and considerations when they wrote introductions, such as why they chose certain data facts as examples. 
Second, the quality control was challenging. 
For example, it was hard to ensure that every participant made enough effort to understand the charts.
\rone{Third, there might be other individual differences among our participants beyond what we measured with our visual literacy survey, such as the amount of domain expertise. Future research could extend our work to include professional participants, such as educators and data journalists, who are experienced in using visualization to communicate with the public, or to collect spoken introductions from publicly available sources. They could also explore alternative approaches to design guidance on reading visualizations, such as conducting co-design workshops~\cite{wang2020cheat} or referencing pedagogy theories~\cite{kwon2016comparative}.}

\textbf{Increase visualization accessibility.}
\rone{The introduction components and structures we crowdsourced could also help design text descriptions of visualizations for visually impaired people to increase visualization accessibility. However, our introductions primarily focused on explaining the visual encoding of a visualization. We would encourage future research to expand this design space beyond just visual encodings, perhaps to include key insights from a visualization or descriptions of how the data was collected for the visualization.}

%% file: section/7-conclusion.tex
\section{Conclusion}
In this paper, we conducted both qualitative and quantitative studies to investigate how to give an effective visualization introduction to a general audience in oral presentations.
We first analyzed 110 introductions written by crowd workers.
These crowdsourced introductions helped us to understand what the general public considered important in a verbal introduction, allowing us to formulate a categorization system that included a diverse set of strategies in visualization introductions. Based on it, we generated different introduction types and evaluated their effectiveness in facilitating visualization comprehension through two extensive experiments with 1,080 subjects.
We compared three ways of giving an explanation (referent, encoding, pattern) and found almost equal effects in facilitating audience comprehension. 
We found that including indicator components (e.g., topic sentences) had negligible effects on improving visualization comprehension, but including knowledge components (e.g., concrete examples) could significantly improve the quality of the introduction and increase understanding above and beyond participants' visual literacy level. 
We hoped that our work could drive more research on visualization presentations and descriptions, making visualizations more accessible for the general public.
% , such as generating more specific guidelines on giving examples and writing topic sentences, and identifying possible sources of chart misunderstanding. 